\documentclass[12pt,a4paper]{article}
\usepackage{graphicx}
\usepackage{times}
\usepackage{url}

\title{\bf Supernovae study: Context of the 4-m ILMT facility}
\author{Brajesh Kumar$^{1,2}$\thanks{brajesh.kumar@iiap.res.in, brajesharies@gmail.com} , S. B. Pandey$^2$,  
        K. L. Pandey$^1$,\\ G. C. Anupama$^1$ and J. Surdej$^3$\\
\vspace{1cm}\\
\normalsize $^1$ Indian Institute of Astrophysics, Koramangala, Bangalore 560 034, India\\ 
\normalsize $^2$ Aryabhatta Research Institute of Observational Sciences, Manora Peak, \\
\normalsize      Nainital 263 001, India \\
\normalsize $^3$ Institut d'Astrophysique et de G\'{e}ophysique, Universit\'{e} de Li\`{e}ge, \\
\normalsize      All\'{e}e du 6 Ao\^{u}t 19, B\^{a}t B5C, 4000 Li\`{e}ge, Belgium \\
}
\date{\mbox{}}
\begin{document}
\maketitle
\pagestyle{empty}
%
%
\def\bull{\vrule height .9ex width .8ex depth -.1ex}
\makeatletter
\def\ps@plain{\let\@mkboth\gobbletwo
\def\@oddhead{}\def\@oddfoot{\hfil\scriptsize\bull\quad
``First Belgo-Indian Network for Astronomy \& astrophysics (BINA) workshop'', held in Nainital (India), 15-18 November 2016 \quad\bull}%
\def\@evenhead{}\let\@evenfoot\@oddfoot}
\makeatother
%
%
\def\beginrefer{\section*{References}%
\begin{quotation}\mbox{}\par}
\def\refer#1\par{{\setlength{\parindent}{-\leftmargin}\indent#1\par}}
\def\endrefer{\end{quotation}}
%
%
{\noindent\small{\bf Abstract:}
The upcoming 4-m International Liquid Mirror Telescope (ILMT) facility will perform deep imaging 
(in single scan $g'$  $\sim$22 mag.) of a narrow strip of sky each clear night in the Time Delayed 
Integration mode. 
A cadence of one day observation will provide unique opportunities to discover different 
types of supernovae (SNe) along with many other types of variable sources. 
We present the approach to discover SNe with the ILMT and discuss the follow-up 
strategy in the context of other existing observational facilities. 
The advantages of the ILMT observations over the traditional glass mirror telescopes are 
also discussed.
}
%
%

\section{Introduction}

The International Liquid Mirror Telescope\footnote{More details about the project
can be found at \url{http://www.ilmt.ulg.ac.be}} (ILMT) will see its first light this year at
the newly developed Devasthal observatory near Nainital (Uttarakhand, India).
It is a 4-m diameter non-conventional telescope with a primary mirror consisting of a 
parabolic shaped rotating container, filled with liquid (mercury).
A detailed description of the site advantages, ILMT components and science cases can be found in 
Surdej et al. (2006), Borra, Hickson and Surdej (2009) and Kumar (2014).
In combination with the $g'$, $r'$ and $i'$ filters, this facility is mainly dedicated to photometric 
and astrometric variability studies along with the detection of supernovae (SNe) as one of the major goals.

Supernova (SN) identification and classification (for a review of different types of SNe see Filippenko 1997)
require the monitoring of light curves and spectra as some of them show transitional features. For example,
Type IIb SNe (Filippenko 1988) closely resemble Type II SNe (presence of hydrogen in the spectra)
during the early phases, while at later epochs this feature disappears and they become more similar to
Type Ib/c events (hydrogen absent).
Therefore, for the proper understanding of these stellar explosions, it is essential to observe them
near their peak brightness and also a follow up at later epochs. It is also very important to detect a SN
at its early phase of explosion as some of the SNe are expected to emit a short burst of high energy
(soft $\gamma$-ray, X-ray, see Nakar \& Sari 2010) radiation at the moment of the shock break-out. 
Thereafter, the cooling will bring the emission into the UV-optical range.
This phase should last at most a few hours, typically less than a day. A dense cadence 
per field would thus allow to systematically detect the shock break-out cooling tail of such events. 
These early observations will be crucial to derive the progenitor radius with a good precision 
(see e.g. Bersten et al. 2013; Taddia et al. 2015).

In the process of detecting SNe, knowing their redshifts, identifying their types, there remain many other 
challenges (see, e.g. Dahlen \& Goobar 2002; Blondin \& Tonry 2007; Kim \& Miquel 2007; Kunz et al. 2007; 
Wang 2007).
There may be a significant level of contamination by other stellar objects (see also
Sect.~\ref{contamination}), for example, Active Galactic Nuclei (AGN\footnote{These are super-massive
black holes in the center of galaxies (Salpeter 1964; Shields 1978)}).
AGNs can be extremely luminous and appear as point sources in imaging surveys. Additionally, they are
situated in the center of their host galaxies and may undergo optical variability (e.g. Stalin et al. 2004).
In a study of the local SN rate, Cappellaro et al. (2005) found a large number of AGNs situated in the
center of their host galaxies. It is possible that they may be mis-identified as SNe in surveys without
spectra and with short observation periods.

It is notable that the ILMT will work in a continuous data acquisition mode by looking only towards the zenith. 
Once a patch of sky has passed over its field of view, it cannot be observed again during the same night. 
Therefore, a collaborative observation will be helpful for the study of transients like SNe. 
Furthermore, the filter system of the ILMT is limited, it will not be sufficient
enough to measure the colour, light curve information. Also, to examine the
spectral features of transients, a spectrum  will be required. Accordingly, large
aperture size traditional mirror telescopes will be needed as complementary to the
ILMT observations.
Thanks to the ARIES observational facilities which presently host three optical telescopes namely the 
1.04-m Sampurnanand Telescope (ST), the 1.3-m Devasthal Fast Optical Telescope (DFOT) and the 
3.6-m Devasthal Optical Telescope (DOT). 
A guaranteed-time allocation strategy to follow-up newly discovered objects will fulfil our needs, 
specially in case of any SN discovery.

\begin{figure}[ht]
\centering
\includegraphics[scale=0.55]{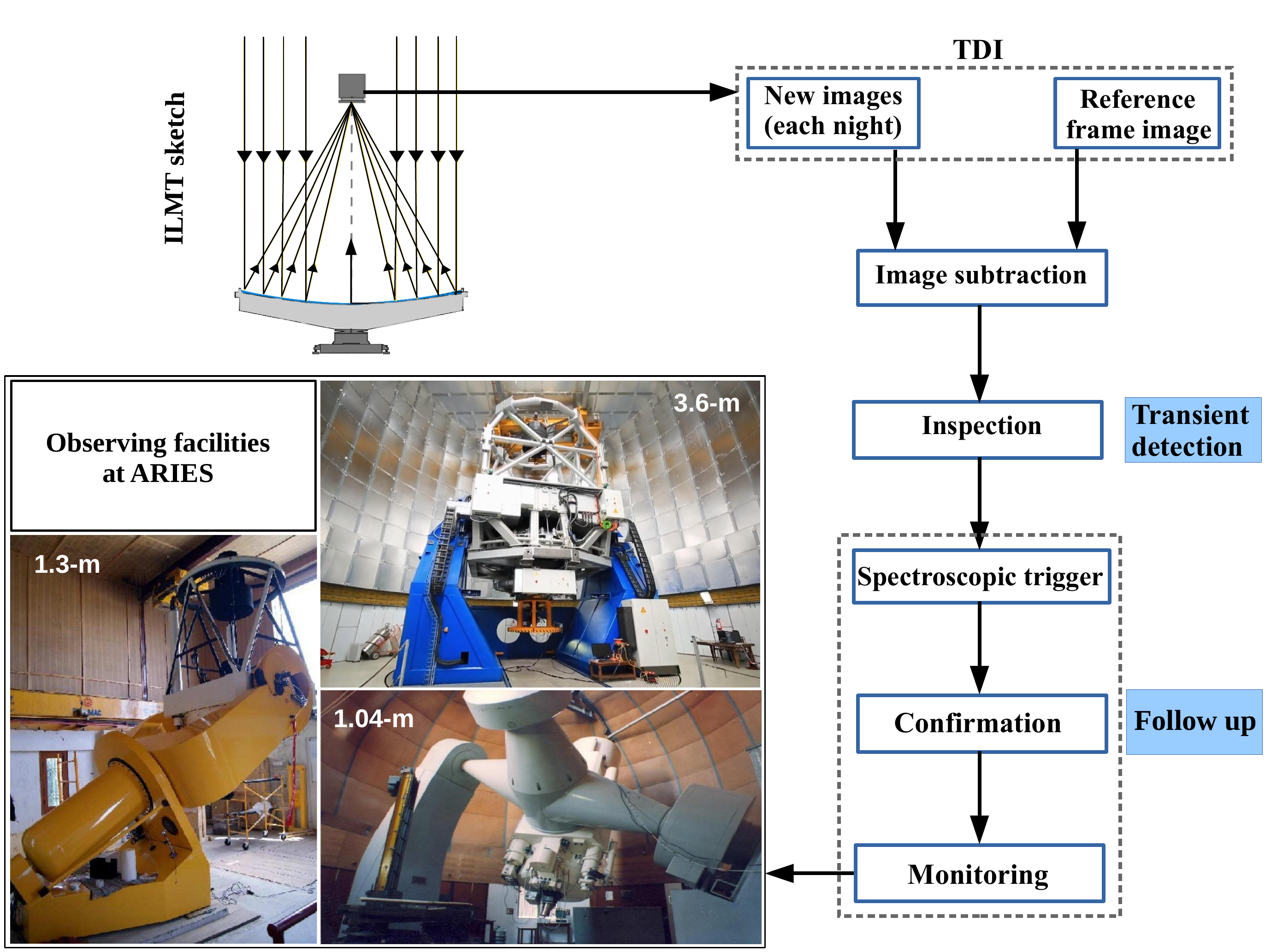}
\caption{Illustration of the proposed processing data flow for SNe detection and
follow-up scheme. Upper left is a sketch of the ILMT and
lower left, images of the existing observing facilities at ARIES, Manora peak and
Devasthal observatories (1.04-m ST, 1.3-m DFOT and 3.6-m DOT).
These telescopes will be used for photometric and/or spectroscopic follow-up 
observations of the ILMT detected SNe and other transient events.}
\label{ilmt_im_sub}
\end{figure}

\section{Supernovae study with the ILMT}

For a reliable and quick source detection in the ILMT images, an automated real time data reduction
pipeline will be developed and applied. In this section we describe the SN detection approach and 
follow-up scheme. A proposed processing data flow is illustrated in Fig.~\ref{ilmt_im_sub}.

\subsection{SN detection approach}

Typically there are two ways for transient detection: i.e. comparison with an exhaustive 
catalog of celestial objects and image subtraction (see also Schmidt 2012). The catalog method
is good where very high precision is required, but it results in poor detection
efficiency near the detection threshold, or in crowded regions. Using the image
subtraction method, images are matched to a template and the template is subtracted.
The latter method is more demanding computationally speaking and has poorer absolute precision,
but leads to a much better transient detection efficiency across a survey.

\subsubsection{TDI mode imaging} \label{tdi}

The ILMT will work in the Time Delayed Integration (TDI) mode (for details see Gibson \& Hickson, 1992; 
Hickson \& Richardson, 1998; Vangeyte et al., 2002, and references therein). There are several 
advantages to work in this mode. As the Earth rotates, the passing stars over the zenith can be imaged 
continuously.  
At the end of the night a single long image of a narrow band sky strip is produced. Although a single 
integration time is imposed however, as the same strip of sky is observed night after night, these 
observations can be co-added to increase the limiting magnitude.
Additionally, TDI imaging also provides an easy and robust way of data reduction. While in conventional 
imaging, the sensitivity irregularities of the CCD sensors are corrected by using a two dimensional flat, 
in TDI mode observations, as the objects go all across the detector along the sensor rows, the sensitivity 
irregularities are averaged over the detector columns. Consequently, the image reduction is done by dividing 
each column by a one-dimensional flat field. Furthermore, this flat field can be directly 
estimated from the scientific data (i.e. the background sky light), contrary to what is done 
during conventional imaging where flat field images must be taken before and/or after 
scientific imaging. In this way precious telescope time is saved. 

\begin{figure*}
\centering
\includegraphics[scale=0.22]{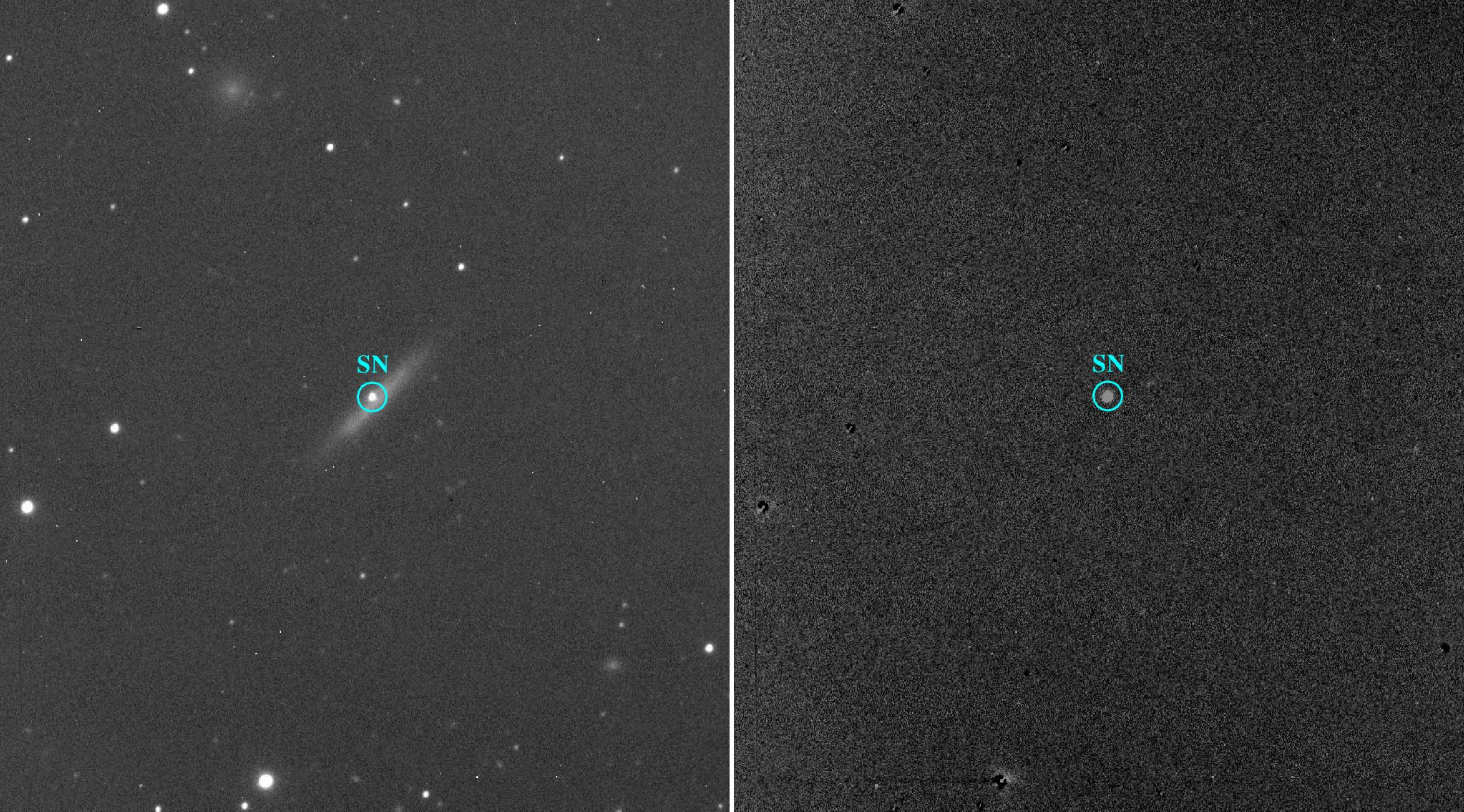}
\caption{Image subtraction. Left panel: Galaxy IC~3311 image with SN~2004gk.
The SN is deeply embedded inside the host galaxy. 
Right panel: Subtracted image where the SN is clearly visible without galaxy 
contamination.}
\label{ilmt_image}
\end{figure*}

\subsubsection{Image subtraction}

Discovering a SN is not an easy task as in most cases the SN light will be a small part 
of light measured from the galaxy. Furthermore, for high redshift galaxies, the galaxies 
themselves will not be fully resolved by ground based observations so a SN will be even 
less distinct and can be easily missed when looking in the individual search epoch images. 
 
In the case of the ILMT, since the same strip of sky will pass over the telescope 
each night, observations will be performed under the best seeing conditions by looking at 
the zenith during each clear night. Then previous night images or a good reference image 
will be subtracted from the 
search night images using the {\it Optimal Image Subtraction} (OIS) technique presented 
in Alard \& Lupton (1998) and further refined in Alard (ISIS, 2000).
This method of image subtraction has already been used to detect SNe in many projects
(e.g. Cappellaro et al. 2005; Wood-Vasey et al. 2007; Poznanski et al. 2007b; Botticella et al.
2008).

Fig.~\ref{ilmt_image} demonstrates one example of the image subtraction technique where both 
images, with and without host galaxy can be seen.


\subsubsection{Transient detection and possible contamination}\label{contamination}

The subtracted images may consist of astrophysical and non-astrophysical sources.
However, in other transient surveys which use image subtraction techniques, it has been found 
that non-astrophysical sources always engulf the real sources (e.g. Bloom et al. 2012; 
Brink et al. 2013).
Non-astrophysical sources include cosmic rays that are not removed by cosmic ray removal 
software, extended features around very bright saturated stars, and bad subtractions due 
to a mis-alignment of the images. To remove these false detections, we must exclude the 
sources around the bright, saturated stars and at the edges of the images. 

Variable stars, quasars, active galaxies and moving objects also contaminate 
the data. We can cross-match the detected object catalog with quasar and AGN
catalogs (V\'eron-Cetty \& V\'eron 2010; P\^aris et al. 2014).
Variable stars can be checked from SIMBAD\footnote{\url{http://simbad.u-strasbg.fr/simbad/}}.
The proper motion of asteroids is large so they will show significant variation
in their position and can be removed easily. Furthermore, solar system 
objects can be checked from the Minor Planet 
Checker\footnote{\url{http://scully.cfa.harvard.edu/cgi-bin/checkmp.cgi}}.

\subsubsection{Detection of supernovae candidates}

It should be highlighted that in general the classification of supernovae is 
done by inspecting the spectra and checking for the presence of emission lines.
But supernovae, especially at high redshift, may be too faint for spectroscopy, 
even with the largest class telescopes currently available. In many cases the 
supernova spectrum is contaminated by the host galaxy light. Furthermore, with the 
increasing number of survey programs, follow-up spectroscopy will not be 
possible/practical for all transients detected in these surveys. At the same 
time unless we confirm that whether a particular event is a core collapse (CC) SN 
or Type Ia, the scientific usefulness will be affected.

Therefore, in response to this need, many techniques targeted at SNe photometric classification have 
been developed which are mostly based on some form of template fitting. These include the methods of 
Poznanski et al. (2002, 2007a); Johnson \& Crotts (2006); Sullivan et al. (2006); Rodney \& Tonry (2009); 
Falck et al. (2010); Gong et al. (2010).
In each of these approaches, typically the light curves in different filters for the SN under 
consideration are compared with those from SNe whose types are well established. Usually, composite 
templates are constructed for each class, using the observed light curves of a number of well-studied, 
high signal-to-noise ratio SNe (see Nugent et al. 2002).  
Some of these light curve fitting models are SALT\footnote{Spectral adaptive light curve template} 
(Guy et al. 2005, 2007), MLCS/MLCS2k2\footnote{Multicolour light-curve shape} (Riess et al. 1995, 
1996; Jha et al. 2007) and SiFTO (Conley et al. 2008).
We must keep in mind that the light curve fitting method is not very robust for typing 
determination and it is less accurate than the spectral method. Nonetheless, it is much easier 
to obtain photometry and construct light curves of faint SNe at high redshift in comparison to 
obtaining spectra (see Melinder 2011). 

Colour information of SNe with the ILMT could be obtained as follows: each night a different filter
($g'$, $r'$ or $i'$) could be used. From the sequence of recorded observations, i.e., $g'$ (day 1), 
$r'$ (day 2), $i'$ (day 3), $g'$ (day 4), $r'$ (day 5), $i'$ (day 6), ... it should be possible to 
interpolate the observed magnitudes on a given date to properly estimate the colour of the SNe on 
that particular day.
For instance the $g'-r'$ colour of the supernova on day 4 would be estimated as 
$g'$(day 4) -- ($r'$(day 2) + ($r'$(day 5) -- $r'$(day 2))*2/3).

For the classification of SNe, Poznanski et al. (2002) presented a method using multicolour 
broadband photometry. Their study is based upon the general assumption that SNe Ic are redder
than SNe Ia at a similar redshift (Riess et al. 2001).
They found that although rising (pre-maximum) SNe Ic have colours similar to those of older 
($\sim$2 weeks past maximum) SNe Ia but near the peak brightness, SNe Ia are typically 0.5 
mag bluer in the $r-i$ colour than SNe Ic. 
Dahlen \& Goobar (2002) and Johnson \& Crotts (2006) also demonstrated similar type determination
methods based on colour cuts and colour evolution.
Bianco et al. 2014 suggested that the colour evolution trends may permit to identify various 
stripped-envelope SNe\footnote{In these events, the outer envelopes of hydrogen and/or helium 
of their progenitors are partially or completely removed before the explosion (e.g. Type IIb, Ib, 
Ic, and Ic-BL).}.
If the ILMT observations could be performed in the subsequent $g'$, $r'$ and $i'$ filters each night, 
the colour information or template fitting technique could be very much useful
to identify the supernova candidates for further follow-up.


\begin{figure*}
\centering
\includegraphics[scale=0.42]{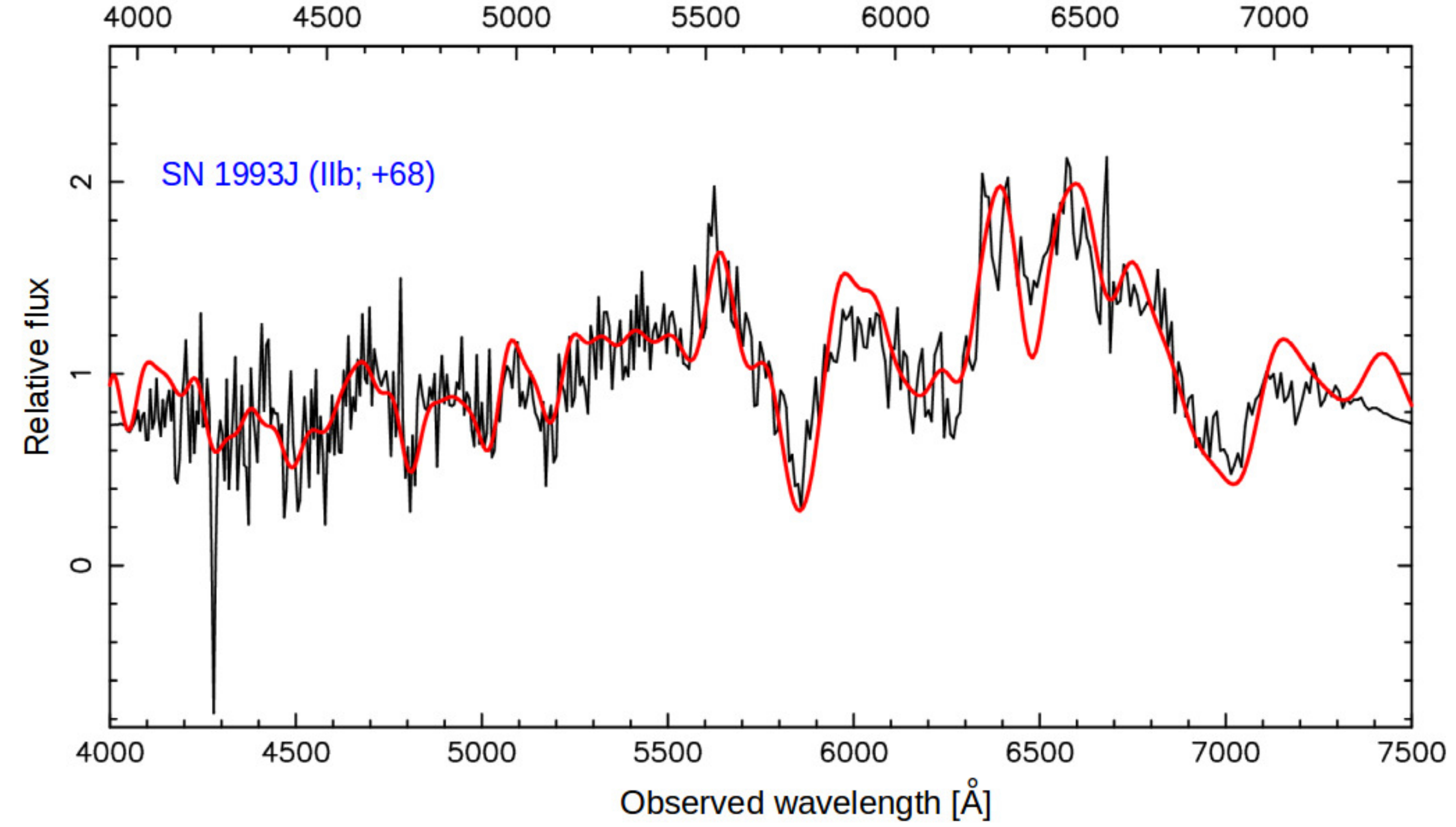}
\caption{Demonstration of the spectra identification using the SNID code. The flux is in arbitrary
units. The observed spectrum of SN~2011fu (Kumar et al. 2013) and template spectrum are, respectively 
shown with black and red colours. The best fitted template is SN~1993J (shown in the top left, blue 
characters with the estimated phase (+68) relative to the maximum light).}
\label{snid_test}
\end{figure*}

\subsection{SN Follow-up scheme}

The light curves, the absolute luminosity and the colour evolution of SNe have provided major insights
to understand their observational properties. Furthermore, the temporal evolution of SNe light curves
constitutes one of the major sources of information about the nature of these events. Therefore, obtaining 
multi-band light curve observations is very important.
Through light curves it has been possible to distinguish between progenitor models, infer some aspects 
of the progenitor evolution, measure the power sources, detail the explosion models, and probe the local 
environment of the supernova explosions (see Leibundgut \& Suntzeff 2003). 
The spectrum of a supernova contains a wealth of information about the composition
and distribution of the elements in the exploding star. It also contains information on
its redshift and age (defined as the number of days from maximum light in a given filter).
It is very important to launch follow-up (photometric and spectroscopic) observations
right after the discovery of any transient.

\subsubsection{Spectroscopic trigger}

Follow-up spectroscopy of a subset of events will be essential to calibrate the accuracy of the photometric 
typing (and host redshifts). In particular, spectroscopy will be invaluable in identifying and quantifying 
catastrophic failures in the typing algorithms; on the basis of these it may be possible to refine the routines.
As SNe are random events, i.e. there is no advance knowledge of their occurrence, 
it is not easy to obtain observing time in advance as in general the time allocation is given 
on the basis of research proposals written several months in advance. Over that successful proposals are
granted only a few nights in a semester on large telescopes. Thanks to the 3.6-m DOT telescope situated near 
the ILMT site, it can be triggered in the target of opportunity observation mode in case of any detected SN 
transient. Additionally, we can collaborate with other existing facilities in India and world wide.


\subsubsection{Confirmation}

The individual emission/absorption lines in the SN spectra provide information about the progenitor 
system and the new elements created in the explosion. Since the classification of SNe is based on 
their optical spectra around maximum light (see Filippenko 1997), therefore, for a rapid confirmation
we can use available SN classification codes e.g. GELATO\footnote{GEneric cLAssification TOol 
\url{https://gelato.tng.iac.es/}} and/or
SNID\footnote{Supernova Identification \url{http://people.lam.fr/blondin.stephane/software/snid/}}.
The SNID tool is developed to determine the type, redshift, and age of a SN, using a single spectrum. 
In Fig.~\ref{snid_test} we show a use of the SNID tool to estimate different parameters of SN~2011fu
(Kumar et al. 2013).
The algorithm is based on the correlation techniques of Tonry \& Davis (1979) and relies on the 
comparison of an input spectrum with a database of high-S/N template spectra (for more details see 
Blondin \& Tonry 2007). 
Furthermore, GELATO is an online software for objective classification of SN spectra. Similar to SNID, 
it performs an automatic comparison of a given (input) spectrum with a set of well-studied SN spectra 
(templates), in order to find the template spectrum that is most similar to the given one. The GELATO 
algorithm is presented in Harutyunyan et al. (2008).

\subsubsection{Monitoring}

Presently there are three optical telescopes existing at ARIES (cf. Fig.~\ref{ilmt_im_sub}).
The 1.04-m ST is situated at Manora peak which is equipped with a modern CCD detector. 
There are two additional telescopes situated at the Devasthal observatory (Sagar et al. 2012), 
the 1.3-m DFOT and 3.6-m DOT. 
Since there are multiple observation facilities available at ARIES, our plan is to quickly 
trigger them once a transient candidate  is confirmed on the ILMT images.
Up to the bright phase ($\sim$17 mag.) of SN, photometric observations will be
performed with different filters using small aperture telescopes (1.04-m and 1.3-m)
and when it will become fainter, larger aperture (3.6-m DOT) telescope will be utilized.
However, spectroscopic observations will be performed with the DOT and other larger 
telescopes in India and/or abroad.

\section{Conclusions}

The ILMT is a unique facility in the context of SNe studies. On each clear night, observation of the same 
sky strip (except for a 4 min. shift in right ascension) will be very effective to apply image subtraction 
and discover new SNe. 
The ILMT will access $\sim$72 square degrees of high galactic latitude ($\mid$b$\mid$ $>$ 30$^\circ$)
in a year. Considering single scan limiting magnitude of $\sim$22 mag ($g'$ filter) and several
other factors (cf. site photometric nights, maintenance etc.), we estimated that it is possible to detect 
hundreds of SNe (both CC and Type Ia) up to moderate red-shifts ($\sim$0.5$z$) every year with this 
telescope. 
However, careful removal of contaminating sources (e.g. cosmic rays, AGNs, variable stars, 
asteroids and minor planets) will be must for SN detection. A detail analysis can be found in Kumar et al. 
(2018).
The SN identification and type determination can be confirmed by employing the light curve and/or 
spectral fitting methods. Once a SN is discovered, rapid monitoring can be performed for more detailed 
investigation. 
In this way, the ILMT imaging will provide an unbiased sample of different types of SNe
which is an advantage over the targeted SN search programs that frequently scan different 
sky regions and possibly induce observational biases. Such a contribution will play a significant 
role to better understand SN physics and stellar evolution. 

%
%
\section*{Acknowledgements}
BK acknowledges the Science and Engineering Research Board (DST, Govt. of India) for financial assistance
in the form of National Post-Doctoral Fellowship (Ref. no. PDF/2016/001563).
JS thanks the Li\`{e}ge University, the F.R.S.-FNRS and the R\'{e}gion Wallonne (Belgium) for their 
constant financial support that has allowed the construction of the 4-m International Liquid Mirror 
Telescope (ILMT). 

%
%

\footnotesize
\beginrefer

\refer Alard C., 2000, A\&AS, 144, 363

\refer Alard C., Lupton R. H., 1998, ApJ, 503, 325

\refer Bersten M. C., Tanaka M., Tominaga N., Benvenuto O. G., Nomoto K., 2013, ApJ, 767, 143

\refer Bianco F.~B., Modjaz M., Hicken M., et al.\ 2014, ApJS, 213, 19

\refer Blondin S., Tonry J. L., 2007, ApJ, 666, 1024

\refer Bloom J.~S., Richards J.~W., Nugent P.~E., et al.\ 2012, PASP, 124, 1175

\refer Borra E., Hickson P., Surdej, J., 2009, Optics and Photonics News, 20, 28

\refer Botticella M.~T., Riello M., Cappellaro E., et al.\ 2008, A\&A, 479, 49

\refer Brink H., Richards J. W., Poznanski D., et al., 2013, MNRAS, 435, 1047

\refer Cappellaro E., Evans R., Turatto M., 1999, A\&A, 351, 459

\refer Cappellaro E., Riello M., Altavilla G., et al.\ 2005, A\&A, 430, 83 

\refer Conley A., Sullivan M., Hsiao E.~Y., et al.\ 2008, ApJ, 681, 482-498 

\refer Dahlen T., Goobar A., 2002, PASP, 114, 284

\refer Falck B. L., Riess A. G., Hlozek R., 2010, ApJ, 723, 398

\refer Filippenko A. V., 1988, AJ, 96, 1941

\refer Filippenko A. V., 1997, ARA\&A, 35, 309

\refer Gibson B. K., Hickson P., 1992, MNRAS, 258, 543 147

\refer Gong Y., Cooray A., Chen X., 2010, ApJ, 709, 1420

\refer Guy J., Astier P., Nobili S., Regnault N., Pain R., 2005, A\&A, 443, 781

\refer Guy J., Astier P., Baumont S., et al.\ 2007, A\&A, 466, 11 

\refer Harutyunyan A. H., et al., 2008, A\&A, 488, 383

\refer Hickson P., Richardson E. H., 1998, PASP, 110, 1081

\refer Jha S., Riess A. G., Kirshner R. P., 2007, ApJ, 659, 122

\refer Johnson B. D., Crotts A. P. S., 2006, AJ, 132, 756

\refer Kim A. G., Miquel R., 2007, Astroparticle Physics, 28, 448

\refer Kumar B., Pandey S.~B., Sahu D.~K., et al.\ 2013, MNRAS, 431, 308

\refer Kumar B., 2014, in PhD Thesis, University of Li\`{e}ge (Belgium)

\refer Kumar B., et al. 2018,  MNRAS, submitted

\refer Kunz M., Bassett B. A., Hlozek R. A., 2007, PhRvD, 75, 103508

\refer Leibundgut B., Suntzeff N.~B.\ 2003, in Supernovae and Gamma-Ray Bursters, 598, 77

\refer Melinder J., 2011, in PhD Thesis, Stockholm University (Sweden)

\refer Nakar E., Sari R., 2010, ApJ, 725, 904

\refer Nugent P., Kim A., Perlmutter S., 2002, PASP, 114, 803

\refer P{\^a}ris I., Petitjean P., Aubourg {\'E}., et al.\ 2014, A\&A, 563, A54

\refer Poznanski D., Gal-Yam A., Maoz D., Filippenko A. V., Leonard D. C.,

\refer Matheson T., 2002, PASP, 114, 833

\refer Poznanski D., Maoz D., Gal-Yam A., 2007a, AJ, 134, 1285

\refer Poznanski D., Maoz D., Yasuda N., et al.\ 2007, MNRAS, 382, 1169 

\refer Riess A. G., Press W. H., Kirshner R. P., 1995, ApJ, 438, L17

\refer Riess A. G., Press W. H., Kirshner R. P., 1996, ApJ, 473, 88

\refer Riess A.~G., Nugent P.~E., Gilliland R.~L., et al.\ 2001, ApJ, 560, 49 

\refer Rodney S. A., Tonry J. L., 2009, ApJ, 707, 1064

\refer Sagar R., Kumar B., Omar A., Joshi Y. C., 2012, ASI Conference Ser., 4, 173

\refer Salpeter E. E., 1964, ApJ, 140, 796

\refer Schmidt B.\ 2012, in New Horizons in Time Domain Astronomy, 285, 9 

\refer Shields G. A., 1978, Nature, 272, 706

\refer Stalin C. S., Gopal Krishna Sagar R., Wiita P. J., 2004, JAA, 25, 1

\refer Sullivan M., et al., 2006, AJ, 131, 960

\refer Surdej J., Absil O., Bartczak P., et al., 2006, in SPIE Conference Series, vol. 6267

\refer Taddia F., Sollerman J., Leloudas G., et al.\ 2015, A\&A, 574, A60 

\refer Tonry J., Davis M., 1979, AJ, 84, 1511

\refer Vangeyte B., Manfroid J., Surdej J., 2002, A\&A, 388, 712 138, 147

\refer V\'eron-Cetty M.-P., V\'eron P., 2010, A\&A, 518, A10

\refer Wang Y., 2007, ApJ, 654, L123

\refer Wood-Vasey W. M., et al., 2007, ApJ, 666, 694

\endrefer           

\end{document}